\documentclass[utf8]{FrontiersinHarvard} 

\usepackage{url,hyperref,lineno,microtype,subcaption}
\usepackage[onehalfspacing]{setspace}

\def\keyFont{\fontsize{8}{11}\helveticabold }
\def\firstAuthorLast{Lattanzi {et~al.}} 
\def\Authors{Valerio Lattanzi\,$^{1,*}$, Miguel Sanz-Novo\,$^{2}$, V\'ictor M. Rivilla\,$^{2}$, Izaskun Jim\'enez-Serra\,$^{2}$, and Paola Caselli\,$^{1}$}
\def\Address{$^{1}$Center for Astrochemical Studies, Max-Planck-Institut f\"{u}r extraterrestrische
Physik, Giessenbachstrasse 1, Garching bei Munchen, 85748, Germany. \\
$^{2}$Centro de Astrobiolog\'{i}a (CAB), CSIC-INTA, Ctra de Torrej\'{o}n a Ajalvir, km
4, Torrej\'{o}n de Ardoz, 28850, Madrid, Spain.}
\def\corrAuthor{Valerio Lattanzi}

\def\corrEmail{lattanzi@mpe.mpg.de}

\begin{document}
\onecolumn
\firstpage{1}

\title[\textit{trans}-HNSO]{Laboratory Detection and Rotational Spectroscopy of \textit{trans}-HNSO: Implications for Astronomical Observations} 

\author[\firstAuthorLast ]{\Authors} 
\address{} 
\correspondance{} 

\extraAuth{}

\maketitle

\begin{abstract}

\section{}

Sulfur-bearing molecules are central to interstellar chemistry, yet their observed abundances in the gas phase remain far below cosmic expectations in dense interstellar regions. Mixed N–S–O species such as thionylimide (HNSO) are particularly relevant, as they incorporate three key biogenic elements. The \textit{cis} conformer of HNSO has recently been detected in the Galactic Center cloud G+0.693-0.027, but no high-resolution data for the higher energy conformer (\textit{trans}-HNSO) had been available until now.

We report the first laboratory detection and rotational spectroscopic characterization of \textit{trans}-HNSO. Spectra were recorded with the Center for Astrochemical Studies Absorption Cell (CASAC) free-space spectrometer employing a hollow-cathode discharge source, yielding 104 assigned transitions between 200 and 530 GHz. A Watson S-reduced Hamiltonian fit reproduced the data with an rms of 40 kHz, providing accurate rotational and centrifugal distortion constants in excellent agreement with CCSD(T) predictions.

Although \textit{trans}-HNSO lies only a few kcal/mol above the \textit{cis} form, it has larger dipole components, making its lines particularly intense (more than 5 times brighter, assuming equal abundances) and a very promising candidate for future astronomical detection. The new measurements enable reliable frequency predictions for astronomical searches and will be added to public databases. Combined with recent evidence for tunneling-driven \textit{trans}-to-\textit{cis} isomerization at cryogenic temperatures, these results open the way to test directly whether quantum tunneling governs the interstellar distribution of HNSO isomers.
\tiny
 \keyFont{ \section{Keywords:} gas-phase chemistry, rotational spectroscopy, sulfur, interstellar: clouds, interstellar: abundances, astrochemistry} 
\end{abstract}

\section{Introduction}

Sulfur chemistry plays a central role in shaping the molecular complexity of the interstellar medium (ISM). Although sulfur is among the ten most abundant elements in the cosmos, with an elemental abundance of S/H $\approx$ 1.3$\times$10$^{-5}$ relative to hydrogen \citep{Asplund2009}, the observed inventory of sulfur-bearing molecules in the gas phase accounts for only a small fraction of its cosmic value. In dense molecular clouds, where the elemental abundances of other species such as carbon, nitrogen, and oxygen are relatively well reproduced by astrochemical models, sulfur appears to be depleted from the gas phase by orders of magnitude compared to the cosmic reference abundance. This long-standing discrepancy is known as the ``missing sulfur problem'' \citep[e.g.][]{laas2019modeling}.\\
Several hypotheses have been proposed to explain this missing sulfur reservoir. On the one hand, it may be sequestered in refractory grain mantles in the form of complex polysulfur chains or sulfur polymers such as S$_8$, which are difficult to observe spectroscopically \citep{Shingledecker2020}. Recent laboratory simulations have demonstrated that hydrogen sulfide (H$_2$S) can be converted on ice-coated interstellar grains in cold molecular clouds through galactic cosmic-ray processing at 5\,K to form sulfanes (H$_2$S$_n$; n = 2–11) and octasulfur (S$_8$) \citep{Herath2025}. This process locks the processed hydrogen sulfide as high-molecular-weight sulfur-containing molecules, providing a plausible rationale for the fate of the missing interstellar sulfur. These sulfur-rich molecules may undergo fractionated sublimation once the molecular cloud transforms into star-forming regions, potentially linking the sulfur chemistry in cold molecular clouds to that in our Solar System. \\
On the other hand, sulfur could be hidden in unstable or elusive molecular carriers that have not yet been identified due to the lack of laboratory spectra, such as sulfur rings and sulfurated polycyclic aromatic hydrocarbon (PASH) molecules \citep{Yang2024}. In this context, the discovery and spectroscopic characterization of new sulfur-bearing molecules are crucial for understanding the distribution of sulfur between gas and dust, constraining chemical models, and enabling targeted astronomical searches with modern observatories.

Beyond this abundance puzzle, sulfur chemistry is deeply intertwined with the chemical evolution of the ISM. Sulfur-bearing molecules participate in both gas-phase and surface reactions, forming diverse functional groups (e.g., –SH, –SO, –CS, and –NS) that act as building blocks for more complex species. Many of these compounds—such as SO, SO$_2$, H$_2$S, OCS, and HCS$^+$—are also sensitive tracers of energetic and dynamical processes, including shocks, photochemistry, and grain-surface desorption, thereby providing key insights into the physical conditions and evolutionary stages of interstellar environments. The richness and reactivity of sulfur chemistry thus contribute significantly to the molecular diversity observed in space and to the formation pathways of prebiotic molecules.
Moreover, sulfur plays a fundamental role beyond the interstellar context. It is a primary product of stellar nucleosynthesis and an important tracer of the chemical evolution of galaxies \citep{Perdigon21}. In planetary atmospheres, sulfur-bearing species drive complex photochemical cycles that influence cloud formation, temperature regulation, and redox balance \citep{Krasnopolsky12,GomezMartin17,Chang23}. On Earth, sulfur is deeply integrated into biochemical processes, including enzymatic activity and protein structure through thiol and disulfide linkages \citep{Richard1986}. More broadly, sulfur has been recognized as one of the essential ingredients for life, bridging planetary, prebiotic, and biological chemistry \citep{Richardson13,Todd22}.

Mixed N–S–O species such as thionylimide (HNSO) are particularly attractive targets for laboratory and astronomical studies. These molecules not only contain sulfur, potentially contributing to the missing sulfur reservoir, but also incorporate nitrogen and oxygen, thus standing out as a relevant molecular link between the interstellar chemistries of these three key prebiotic elements. On Earth, several HNSO isomers have been suggested to connect the biochemistries of nitric oxide (NO) and hydrogen sulfide (H$_2$S), two fundamental gasotransmitters in living beings, further strengthening their astrobiological relevance \citep{Filipovic2012,Miljkovic2013,Ivanova2014,Wu2018,Zhao2024}. In this context, the laboratory characterization of the rotational spectrum of HNSO is crucial to provide accurate rest frequencies for astronomical searches, particularly in environments rich in sulfur-bearing molecules. Additionally, understanding the stability and relative abundance of its isomers can shed light on the chemical pathways that transform sulfur into different gas-phase and solid-phase carriers, thereby bridging laboratory studies with astronomical observations.

The HNSO system exists as a small family of structural isomers whose relative energies and interconversion barriers critically determine which forms may be present under interstellar conditions. High-level quantum-chemical work has shown that several [N,S,O]-containing isomers (including HNSO and related SNO/NSO species) are close enough in energy that both formation and isomerization are plausible in gas-phase and surface chemistries relevant to the ISM \citep{Kumar2017}. Recent advances in composite quantum-chemical strategies and spectroscopic predictions (which integrate CCSD(T)-quality energetics with anharmonic and core-valence corrections) have further improved the reliability of computed geometries, rotational constants and isomerization barriers for NSO/SNO-type moieties, making such calculations an increasingly powerful complement to laboratory work \citep{Barone2024}. Experimentally, the rotational and rovibrational spectroscopy of thionylimide (HNSO) has a long history, but has been concentrated on the \textit{cis} conformer: the earliest microwave assignment to planar \textit{cis}-HNSO was reported in the late 1960s, and subsequent laboratory studies extended and refined those microwave measurements and the centrifugal-distortion analysis \citep{Kirchhoff1969,DalBorgo1979}. Cavity Fourier transform microwave (FT-MW) work later resolved nuclear hyperfine structure in HNSO and its isotopologues, providing precise $^{14}$N and D hyperfine parameters \citep{Heineking1993}. High-resolution infrared studies have also characterized several strongly perturbed vibrational bands of HNSO, supplying valuable vibrational–rotational constants and aiding spectroscopic modeling \citep{Puskar2006}.

By contrast, a comprehensive, high-resolution rotational characterization of the \textit{trans} conformer remains absent in the literature. While quantum-chemical calculations indicate that the \textit{trans} form is not more stable than the \textit{cis} conformer, it lies less than 4\,kcal/mol higher in energy \citep{Kumar2017,Barone2024}, suggesting that it could still be appreciably populated under interstellar conditions. Importantly, the \textit{trans} isomer exhibits larger dipole moment components along both the $a-$ and $b-$axes compared to the \textit{cis} form, enhancing the intensity of its rotational transitions and making it potentially more detectable in the gas phase despite its slightly higher energy. This is particularly relevant in light of the recent detection of \textit{cis}-HNSO toward the Galactic Center cloud G+0.693-0.027 (G+0.693 hereafter) by \citet{SanzNovo2024}, where the emission lines were sufficiently intense to derive a significant abundance for the \textit{cis} conformer ($\sim$\,4.8 times lower than SO$_2$). 

The detection of high-energy isomers in the ISM has traditionally been viewed with skepticism due to their expected low abundances compared to more stable counterparts \citep{Lattelais2009}. However, recent radioastronomical observations have challenged this view, revealing that several molecular systems contain either entire families of structural isomers or selectively the higher-energy members (e.g. C$_3$H$_2$O, C$_2$H$_4$O$_2$, C$_2$H$_2$N$_2$, C$_2$H$_5$O$_2$N, and HOCS$^+$; \cite{Shingledecker2019,Mininni2020,Rivilla2023,SanAndres2024,SanzNovo2024b}). For instance, in the HSCO$^+$/HOCS$^+$ pair ($\Delta$E = 4.9\,kcal/mol; \citet{Wheeler2006}), only the higher-energy isomer HOCS$^+$ has been detected \citep{SanzNovo2024b}, while within the C$_2$H$_5$O$_2$N family, only glycolamide is observed—despite lying 9.7\,kcal/mol above the global minimum (N-methylcarbamic acid). These observational results imply that even a higher-energy isomer like \textit{trans}-HNSO, especially with its stronger dipole, could produce detectable emission. The lack of laboratory rotational data suitable for precise astronomical rest-frequency predictions, however, has so far limited targeted searches and incorporation into spectroscopic databases such as CDMS\footnote{https://cdms.astro.uni-koeln.de/} and JPL\footnote{https://spec.jpl.nasa.gov/} which hinders its interstellar search, further prompting new dedicated experimental effort. 

The present study reports the first laboratory detection and high-resolution rotational spectroscopic characterization of \textit{trans}-HNSO. Using frequency-modulated absorption spectroscopy in the microwave and millimeter-wave regions, we have assigned and analyzed a set of transitions that allow us to determine accurate spectroscopic parameters, including rotational and centrifugal distortion constants. These data provide the foundation for reliable spectral predictions over a wide frequency range, enabling future astronomical searches. In addition, we discuss the astrochemical implications of our findings, particularly regarding the detectability of \textit{trans}-HNSO in sulfur-rich regions of the ISM and its relationship to the \textit{cis} conformer.

\section{Experimental methods}

The measurements were carried out with the Center for Astrochemical Studies Absorption Cell (CASAC), a frequency-modulated free-space absorption spectrometer developed at the Max-Planck Institute for Extraterrestrial Physics \citep{Bizzocchi2017}. CASAC has been used extensively to characterize reactive sulfur-bearing species and other \textit{trans}ient molecules \citep[e.g.][]{Prudenzano2018,Lattanzi2018,Inostroza-Pino2023,Araki2024,Lattanzi2024}, and the present experiments follow the same detection philosophy and instrumental approach.

The primary radiation source was a Keysight E8257D frequency synthesizer phase-locked to a 10\,MHz rubidium frequency standard (Stanford Research Systems), providing high absolute frequency accuracy and phase stability. The synthesizer output was amplified and multiplied by Virginia Diodes solid-state active multiplier chains to cover the millimeter and submillimetre bands (instrumental capability $\approx$\,75–1600\,GHz). Frequency modulation (FM) was applied at 50\,kHz, and the FM signal after interaction with the plasma was detected by a cryogen-free InSb hot-electron bolometer (QMC Instruments Ltd.). The detector output was demodulated with an SR830 lock-in amplifier (Stanford Research Systems) at the second harmonic (2\textit{f}), producing the characteristic second-derivative line shapes of FM absorption spectroscopy. Instrument control, data acquisition and averaging were automated via a computer-controlled acquisition system.

Radiation traversed a 3.0\,m long, 5\,cm diameter Pyrex absorption tube. For these measurements we employed the hollow-cathode discharge source recently developed in our laboratory and mounted inside the Pyrex tube to create the plasma region used for species production. The hollow-cathode arrangement proved effective at producing and enhancing the concentration of N–S–O reaction products and was used for all reported spectra. A mixture of SO$_2$ and NH$_3$ gas in a 1:3 ratio diluted in argon was introduced to the cell, yielding a downstream (measured) total pressure of $\sim$\,20\,mTorr. The discharge was operated in hollow-cathode mode at 600\,V and 85\,mA, conditions optimized to maximize the production of the known \textit{cis} isomer while maintaining stable plasma operation. The cell was kept around freezing temperature with a tiny amount of liquid nitrogen to reduce Doppler widths and avoid overheating of the system due to the discharge.

\begin{figure*}[htbp]
    \centering
    \includegraphics[scale=0.13]{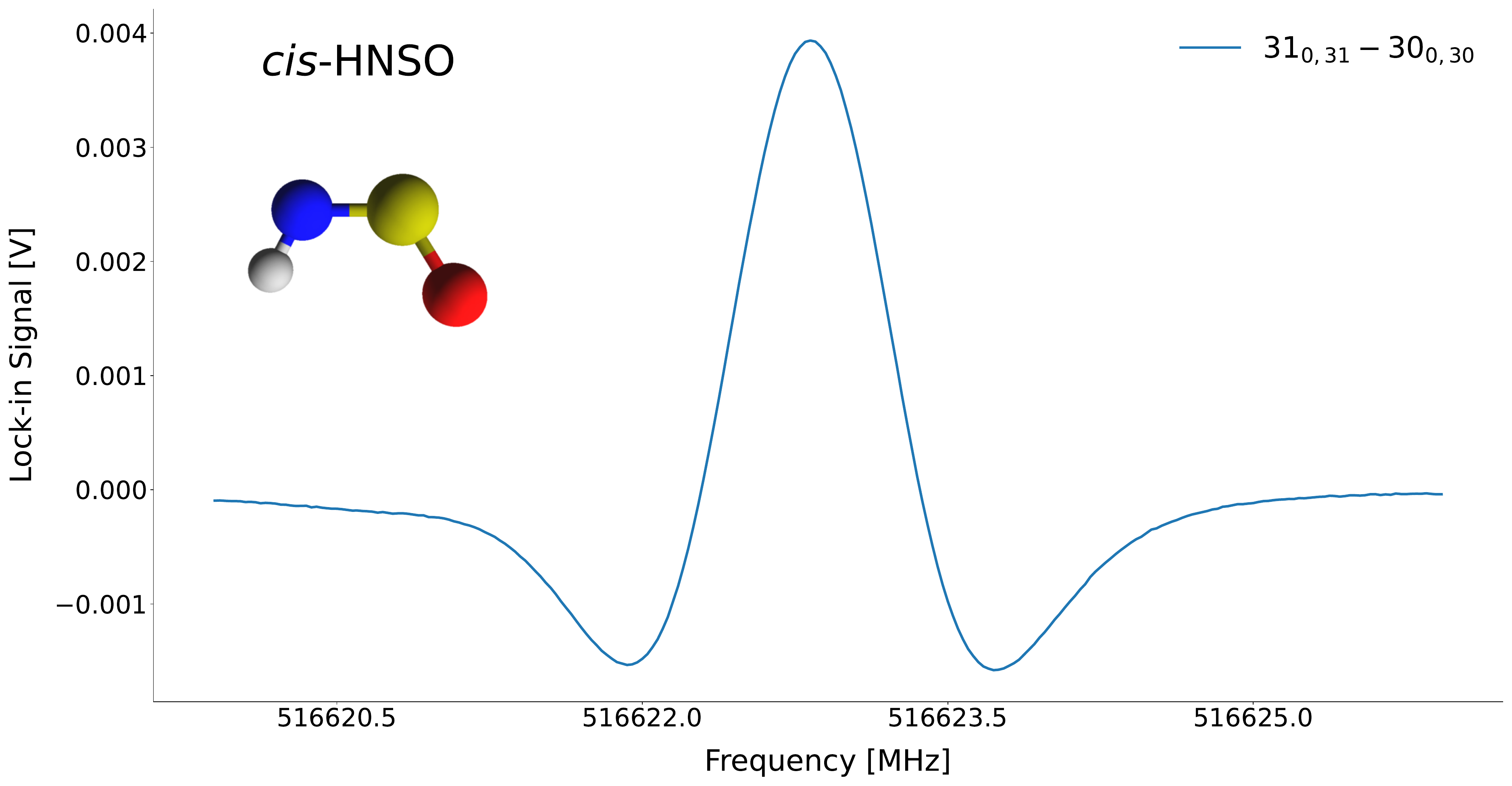}
    \includegraphics[scale=0.13]{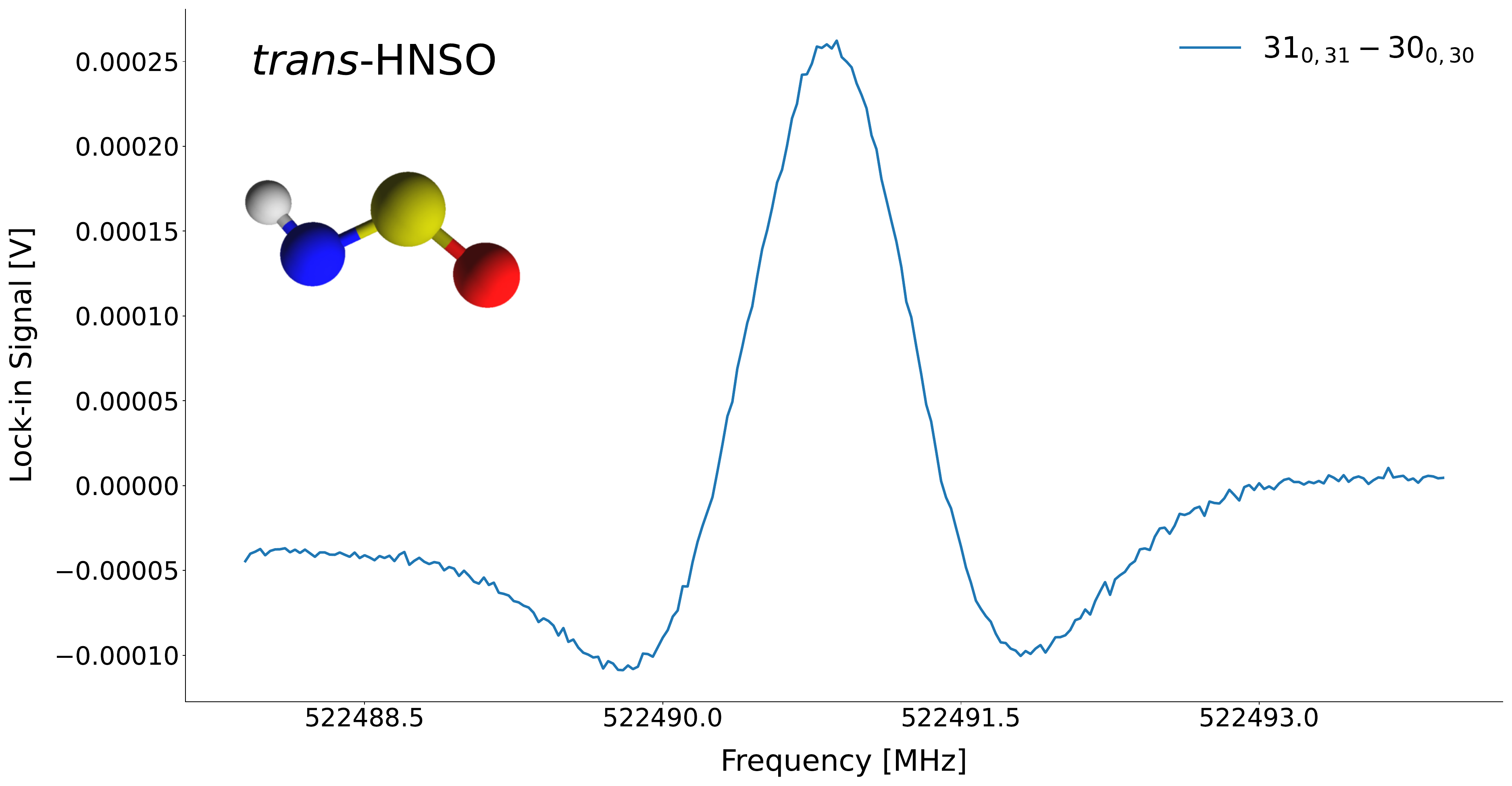}
    \caption{Raw experimental spectra of \textit{cis}- (left) and \textit{trans}-HNSO (right); both spectra were acquired with a 3\,ms time constants and an integration time of 20 and 180 seconds for \textit{cis} and \textit{trans} respectively. Although the y-axis labels are arbitrary, they have been retained to highlight the differences in signal intensity between the two isomers when measured under identical experimental conditions.} 
    \label{fig:spectra}
\end{figure*}

The calibration and initial searches benefited from very strong \textit{cis}-HNSO calibration lines (e.g. Fig.\,\ref{fig:spectra}), which were observed with extremely high intensity and required essentially no extensive exploration of experimental degrees of freedom. These robust \textit{cis} signals were used to verify instrument performance and coarse frequency placement. Upon switching focus to the \textit{trans} conformer, diagnostic scans showed that \textit{trans} lines were detectable after only a few scans; this allowed rapid optimization of the hollow-cathode operating point to the final conditions reported above. Under these optimized settings, individual transitions of the \textit{trans} conformer were observed with good signal-to-noise using minimal averaging, and spectra remained stable over extended acquisitions.

Because many reactive-species experiments can suffer from intrinsically weak absorptions, it is worth noting that the favorable signal strength and stability obtained here substantially simplified line assignment and reduced systematic uncertainty in frequency determination.

All frequencies reported in this work were measured under the operating conditions given above. Uncertainties reflect combined contributions from line fitting and signal to noise quality; details of the line fitting, Hamiltonian model and uncertainty treatment are given in the Analysis subsection below.

\begin{table*}[htbp]
\caption{Spectroscopic parameters of \textit{trans}-HNSO.\\ Values in parentheses represent 1$\sigma$ uncertainties, expressed in units of the last quoted digit.\\
$^a$Rotational constants were derived from the equilibrium values obtained at the CCSD[T]/cc-pwCVQZ level and corrected for the vibrational contributions ($B_e-B_0$) estimated at CCSD[T]/cc-pV(T+d)Z. Centrifugal distortion terms derived from anharmonic force field calculations at
CCSD(T)/cc-pV(T+d)Z. \\
$^b$From \citet{Barone2024}.\\
$^c$Fixed to the theoretical value.\\
$^d$Dimensionless rms, defined as $\sigma_{w} = \sqrt{\frac{\sum_i\left(\delta_i/err_i\right)^2}{N}}$, where the $\delta$'s are the residuals weighted by the experimental uncertainty (\emph{err}) and \emph{N} the total number of transitions analyzed.}
\label{table:1} \centering
\begin{tabular}{llr@{.}lr@{.}lr@{.}lr@{.}lr@{.}l}
\hline\hline   
\noalign{\smallskip}
Parameter & unit & \multicolumn{2}{c}{Experimental} & \multicolumn{2}{c}{Theory (this work)$^a$} & \multicolumn{2}{c}{Theory (previous)$^b$} \\
\hline
\noalign{\smallskip}
   $A_0$       & MHz  &      50429&061(23)        &  50848&857      &  50740&9     \\
   $B_0$       & MHz  &       9957&14715(77)      &   9950&057      &   9962&9     \\
   $C_0$       & MHz  &       8297&86728(70)      &   8304&240      &   8310&3     \\
\noalign{\smallskip}
\hline
\noalign{\smallskip}
   $D_J$     & kHz  &          5&6815(13)       &      5&56824      &  \multicolumn{2}{c}{}    \\
   $D_{JK}$  & kHz  &       --63&432(54)        &   --62&7970       &  \multicolumn{2}{c}{}    \\
   $D_K$     & MHz  &          1&3911(31)       &      1&33535      &  \multicolumn{2}{c}{}    \\
   $d_1$     & kHz  &        --1&51879(24)      &    --1&48506      &  \multicolumn{2}{c}{}    \\
   $d_2$     & kHz  &        --0&09954(23)      &    --0&0880475    &  \multicolumn{2}{c}{}    \\
   $H_J$     & Hz   &          0&00159(73)      &      0&0072804    &  \multicolumn{2}{c}{}    \\
   $H_{JK}$  & Hz   &        --0&2631562$^{c}$  &    --0&2631562    &  \multicolumn{2}{c}{}    \\
   $H_{KJ}$  & Hz   &        --10&9(14)         &    --6&0255433    &  \multicolumn{2}{c}{}    \\
\hline
\noalign{\smallskip}
\# lines       &      & \multicolumn{2}{c}{104}   & \multicolumn{2}{c}{} & \multicolumn{2}{c}{} \\
$\sigma_{rms}$ &  kHz & \multicolumn{2}{c}{40}    & \multicolumn{2}{c}{} & \multicolumn{2}{c}{} \\  
$\sigma_{w}^c$ &      & \multicolumn{2}{c}{0.79}  & \multicolumn{2}{c}{} & \multicolumn{2}{c}{} \\
\hline
\end{tabular}
\end{table*}

\section{Theoretical calculations}

All quantum-chemical calculations were performed with CFOUR (Coupled-Cluster techniques for Computational Chemistry) \citep{matthews2020coupled} using the CCSD(T) model \citep[coupled-cluster with single and double excitations and a perturbative treatment of triple excitations,][]{Raghavachari1989}. 
Geometries were obtained in the frozen-core approach with cc-pV(X+d)Z (with X\,=\,T,Q,5) Dunning’s correlation-consistent polarized valence basis sets of zeta quality with additional tight d-type polarization function added to sulfur \citep{DunningPetersonWilson2001} to increase the accuracy of the calculated geometries. When considering all electrons in the correlation treatment, the cc-pwCVXZ (X\,=\,T and Q) basis sets \citep{Peterson2002} were used. The best equilibrium structures of \textit{trans}-HNSO have been calculated at the CCSD[T]/cc-pwCVQZ level of theory (Table\,\ref{table:1}), which has been shown on several occasions to yield equilibrium structures of very high quality for molecules harboring second-row elements (e.g. \cite{Lattanzi2010}). Centrifugal distortion constants were evaluated from anharmonic force-field calculations with second-order vibrational perturbation theory (VPT2) at the CCSD(T)/cc-pV(T+d)Z level.

\section{Analysis}

HNSO exists in \textit{cis} and \textit{trans} conformers with \textit{cis} being 3.7\,kcal/mol ($\sim$\,1860\,K) relatively more stable (see Figure\,S1), and a \textit{cis}-\textit{trans} conversion barrier of 14.5 kcal/mol (at the CCSD(T)/aug-cc-pV5Z//CCSD(T)/aug-cc-pVTZ level; \citealt{Kumar2017}). The \textit{trans}-HNSO is an asymmetric-top rotor very close to the prolate limit ($\kappa=-0.92$) with a $C_s$ symmetry and a quasi planar structure (inertial defect of $\Delta \approx 0.128$\,amu\,$\AA^2$). The present rotational analysis is based on a dataset of 104 distinct microwave transitions fitted with 11 spectroscopic parameters (see Table\,\ref{table:1}), covering the frequency interval $\sim$\,200-530\,GHz and sampling upper-state quantum numbers up to $J=31$  and $K_a$ up to 6. This combination of line density, $J$-range and $K_a$-coverage gives good leverage on all three principal rotational constants and on the leading centrifugal-distortion terms required to model the measured molecule. To determine the central frequencies of the observed transitions, each averaged experimental spectrum was first baseline-corrected and then fitted using a modulated Voigt line-shape function \citep{Dore2003}. The fitting procedure was carried out with QtFit, a routine included in our in-house Python library for laboratory spectroscopy (pyLabSpec\footnote{https://laasworld.de/pylabspec.php}). In order to reproduce subtle line-shape effects, the analysis also accounted for both the dispersive component of the Fourier transform of the dipole correlation function and a low-order polynomial term (typically second or third order). These additional contributions compensate for line asymmetry and residual baseline structure caused by standing waves within the absorption cell due to imperfect transmission of the window materials. 

Fitting the observed transitions with a Watson S-reduced Hamiltonian yielded an overall standard deviation of 40\,kHz (see Table\,\ref{table:1}). Individual transitions uncertainties were set to 50\,kHz for the majority of lines, while three particularly weak transitions were assigned larger uncertainties of 100\,kHz. The normalized rms (reduced $\chi^2$ equivalent) is 0.79, i.e. very close to unity, which implies that the adopted experimental uncertainties and the Hamiltonian model are mutually consistent and that no overall inflation of the uncertainties is required. 

The quantum-number coverage makes the dataset sensitive to the three principal rotational constants ($A, B, C$) and to both diagonal and off-diagonal centrifugal distortions. In addition, earlier theory predicts appreciable $\mu_a-$ and $\mu_b-$dipole moment components for the \textit{trans} conformer (1.96\,D and 2.41\,D derived from our calculations at CCSD(T)/cc-pwCVQZ level). No obvious systematic residual structure attributable to partially resolved nuclear hyperfine splitting is apparent at the present resolution, linewidth, and RMS level, considering also the quantum number coverage of the present dataset ($J_{min}=10$) and the collapse of the hyperfine structure with higher $J$'s. The hyperfine structure from $^{14}$N might be discoverable by a dedicated search in the radio band in our molecular jet expansion system coupled with our chirped-pulse spectrometer (CASJet+CP-FT).

The rotational constants are determined experimentally with very high precision (uncertainties from a few 10$^{-2}$ MHz down to 10$^{-5}$ MHz). The theoretical values at the quoted level (CCSD[T]/cc-pwCVQZ and corrected for the vibrational contribution estimated at CCSD[T]/cc-pV(T+d)Z), are in good agreement, especially for $B$ and $C$ with differences that are small in absolute terms ($B$: 17\,MHz, $C$: 7\,MHz), and correspond to fractional deviations of $\leq$\,0.07\,\%. These small differences are consistent with expectations for the level of the quantum chemical calculated geometries that do not include full vibrational averaging and complete-basis extrapolation. The excellent match confirms the conformational assignment and indicates that the calculated geometry captures the heavy-atom framework with high fidelity.
Overall also the quartic distortion constants agree to within a few percent — a level of correspondence that is satisfactory given that centrifugal constants are particularly sensitive to the chosen level of theory (harmonic force field vs. full anharmonic treatment) and to vibrational corrections. The relatively larger fractional discrepancy seen for the smallest parameters (e.g., $d_2$) is expected because when the absolute magnitude is tiny even small absolute differences produce large relative percentages; in practice the absolute differences here are only a few $10^{-3}-10^{-2}$\,kHz.

Only three sextic parameters were used in the model to reproduce the experimental lines within the estimated uncertainties. One of these, namely $H_{JK}$, was kept fixed to the ab initio calculated value since the frequency/quantum number coverage was not able to efficiently constrain it, although removing it completely worsened the overall fit. The remaining sextic parameters were marginally determined from the data and, although their formal uncertainties are larger than those of the rotational and quartic distortion constants, their absolute magnitudes are small and their inclusion materially improves the reproduction of the high-frequency lines within the measured range. In particular, the experimental values for $H_J$ and $H_{KJ}$ provide useful empirical anchors that, together with the calculated values, enable confident predictions of transition frequencies across the observed band.

\section{Discussion and Conclusions}

The present study reports the first high-resolution rotational characterization of \textit{trans}-HNSO, establishing a robust spectroscopic foundation for its astronomical detection and placing the molecule in a broader astrochemical context. The results provide both accurate rest frequencies and a deeper understanding of how this conformer compares with theoretical expectations, while also highlighting the implications of its potential presence in space. Our dataset yields rotational and centrifugal distortion constants of remarkable precision, with an overall rms deviation of 40\,kHz, and a general good agreement with the couple-cluster predicted values. By combining experimentally determined constants with selected theoretical values where necessary, we obtain a balanced set of parameters that provides highly reliable frequency predictions within the measured region, while maintaining controlled extrapolation beyond it.

\begin{figure*}[htbp]
    \centering
    \includegraphics[scale=0.35]{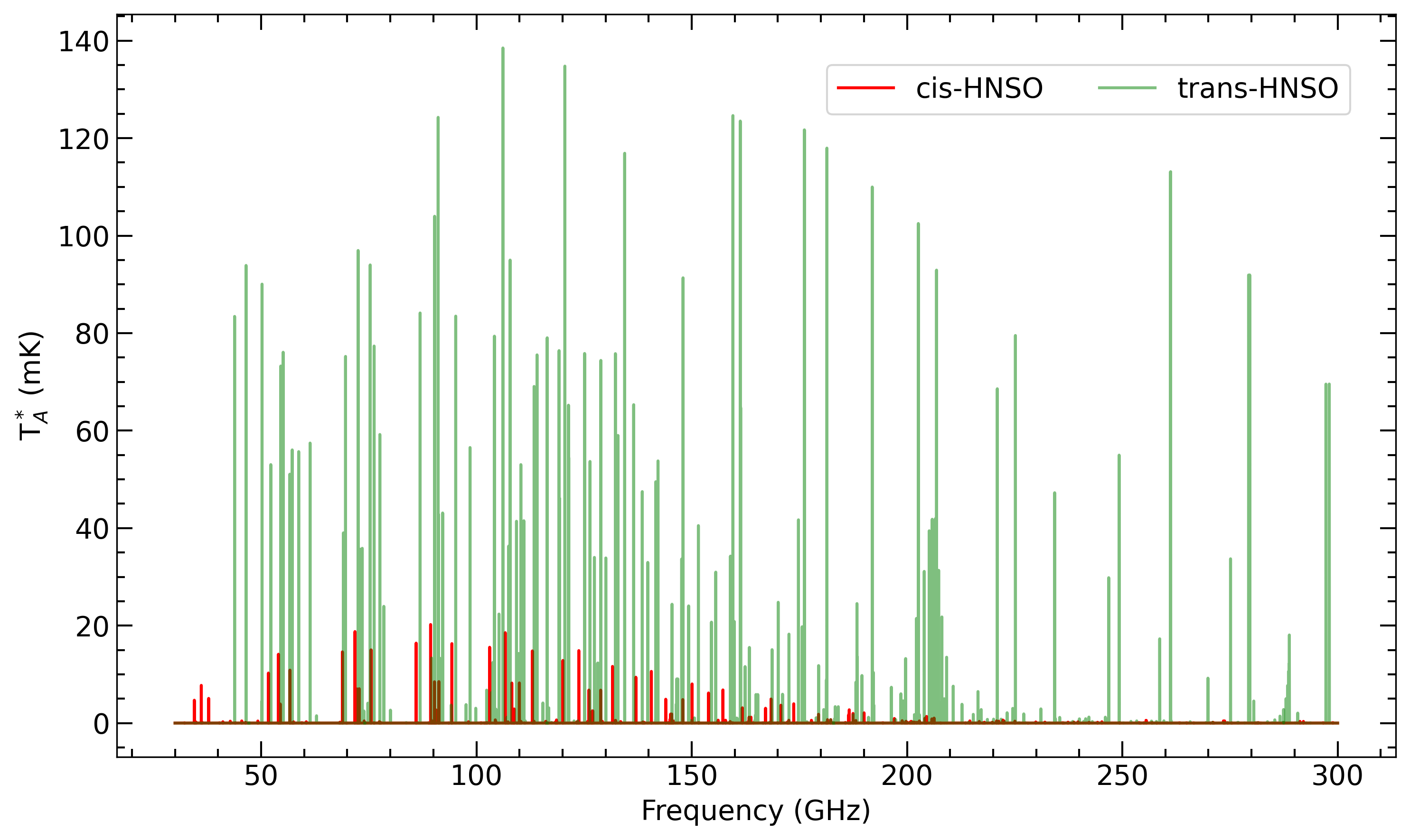}
    \includegraphics[scale=0.35]{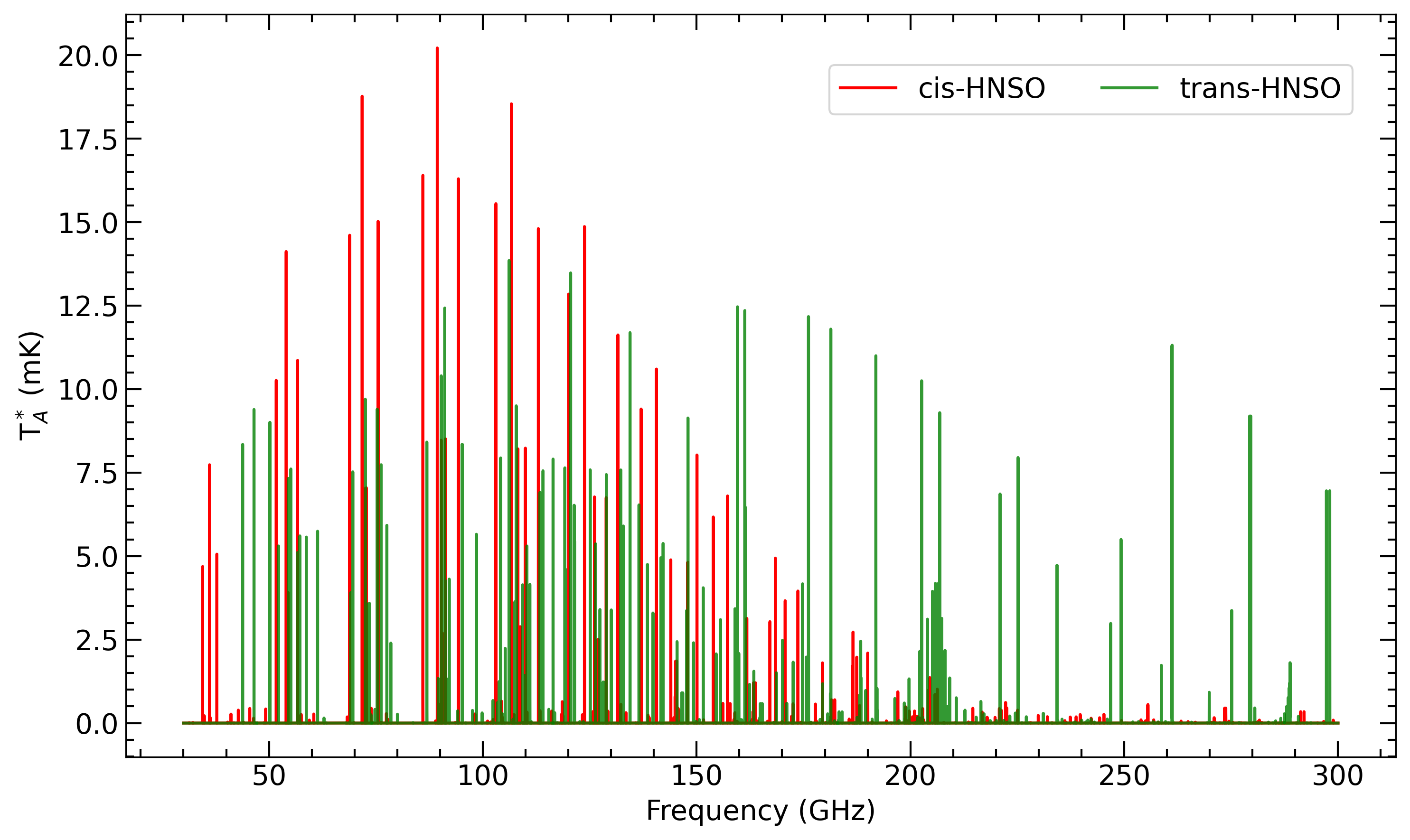}
    \caption{Simulated spectra of the two HNSO conformers generated using the MADCUBA-SLIM tool \citep{Martin2019} and assuming (\textit{left}) the same physical parameters derived for $cis$-HNSO toward G+0.693 (i.e., $N= 8\times10^{13}$cm$^{-2}$, $T_{ex}$\,=\,11\,K and FWHM\,=\,21.5\,km/s; \citealt{SanzNovo2024}), (\textit{right}) a 10:1 abundance ratio of $cis/trans$.} 
    \label{fig:astro_predictions}
\end{figure*}

Two features make \textit{trans}-HNSO a promising target for radioastronomical searches. First, its dipole moment has substantial components along both the a- and b-axes, ensuring that a wide range of transitions exhibit significant intensities. Second, the present frequency determinations are anchored with sub-50 kHz accuracy, far surpassing the resolution requirements imposed by typical interstellar line widths, such as the case of G+0.693 with typical linewidths of about 15-20\,km/s ($\sim$\,5-7\,MHz at 100\,GHz; \citealt{requena-torres_lar_2008,zeng2018}). These qualities suggest that, provided it reaches detectable abundances, \textit{trans}-HNSO can be securely identified in observational spectra (see Figure\,\ref{fig:astro_predictions}). With an estimated frequency accuracy on the order of a few tens of kHz for the strongest transitions in the Q band (33-50 GHz) and the 3\,mm band ($\sim$\,80–120\,GHz), the spectroscopic database derived from our new measurements provides a reliable basis for the identification of this conformer also in cold Galactic environments, such as TMC-1, where molecular emission lines typically exhibit linewidths of ~0.5\,km/s (~167\,kHz).

.

This motivation is strengthened by the recent detection of the \textit{cis} isomer in the Galactic Center cloud G+0.693 (Sanz-Novo et al. 2024), where its lines were strong enough to permit abundance estimates. Moreover, G+0.693 has already emerged as a uniquely rich source of molecular stereochemistry, with several distinct stereoisomers detected toward this cloud (e.g. the Z- and E- conformers of cyanomethanimine \citep{Rivilla2019}, \textit{cis} and \textit{trans} HCOOH \citep{SanzNovo2023}, n-propanol n-C$_3$H$_7$OH \citep[n-C$_3$H$_7$OH; ][]{Jimenez-Serra2022}, the high-energy \textit{trans} conformer of carbonic acid, HOCOOH \citep{SanzNovo2023},  \textit{trans} methyl formate \citep{SanzNovo2024} and \textit{cis} N-methylformamide \citep{Zeng2025}. The coexistence of multiple isomeric forms in the same environment indicates that both formation and survival of higher-energy isomers are feasible under specific interstellar conditions. Given that the \textit{trans} conformer is only a few kcal/mol less stable than \textit{cis}-HNSO, but possesses larger dipole moments, it may therefore be present with the current levels of sensitivity of the above ultra-deep survey. Sources such as G+0.693, Sgr\,B2(N), Orion\,KL, and shocked outflows thus represent natural starting points for targeted searches. To support such efforts, the spectroscopic parameters derived here will be incorporated into public databases, such as CDMS.

The simultaneous availability of \textit{cis} and \textit{trans} spectroscopic data opens new avenues for using their relative abundances as probes of interstellar chemistry. A dominance of \textit{cis}-HNSO would point to low-temperature formation pathways or grain-surface synthesis with limited energy available for isomerization. By contrast, the detection of \textit{trans}-HNSO at appreciable levels would indicate either gas-phase routes capable of producing the higher-energy conformer or in general other competitive chemical processes that promote isomerization in the gas phase \citep[as e.g. multidimensional small curvature tunneling effects;][]{Concepcion2022}. Quantifying the \textit{cis}/\textit{trans} ratio in different environments would therefore provide direct constraints on the dominant chemical routes of sulfur–nitrogen–oxygen chemistry in the ISM \citep[see also ][]{Molpeceres2025}.

Since extrapolation toward lower frequencies is generally reliable when higher-frequency measurements are available, predictions up to the measured range should be regarded as sufficiently accurate for radioastronomical identification. However, further improvements might be desirable to refine the quality of extrapolations when higher frequency transitions are needed. Extending the measurements to higher frequencies and higher $J$ levels would provide tighter constraints on sextic and higher-order distortion parameters. Complementary isotopologue studies could deliver independent structural constraints and reduce correlations among centrifugal constants. In this context, the relatively high abundance observed for \textit{cis}-HNSO makes its $^{34}$S isotopologue a particularly promising target, motivating new dedicated laboratory measurements.
Moreover, low frequency, i.e. radio/microwave spectroscopy could probe the lowest-frequency transitions with sub-kHz precision, potentially resolving hyperfine structure and thereby enhancing catalog accuracy. Parallel theoretical work, including vibrational corrections and higher-level treatments of electron correlation, will also help to benchmark the most delicate parameters.

An intriguing additional dimension to the problem of HNSO isomerism has recently been provided by \citet{Jian2025}, who demonstrated that \textit{trans}-HNSO generated photochemically in cryogenic matrices undergoes rapid conversion back to the \textit{cis} form via quantum tunneling, with a half-life of only minutes at 3 K. While these results clearly indicate efficient tunneling-driven isomerization under matrix-isolation conditions, it remains uncertain to what extent such behavior would persist in the gas phase or in astrophysical environments. The precise spectroscopic constants reported here now enable this hypothesis to be tested observationally. If \textit{trans}-HNSO is detected in regions where \textit{cis}-HNSO is abundant, it would indicate that tunneling is less effective under interstellar conditions than in laboratory ices, perhaps due to environmental differences or competing processes, and in the gas phase, as suggested by \citet{Concepcion2021} to explain the E/Z isomer ratio of several imines detected in the ISM. Conversely, if \textit{trans}-HNSO remains systematically undetected despite favorable conditions, it would strongly support the view that tunneling controls its astrochemical fate. In either outcome, astronomical observations made possible by this work will provide an unprecedented opportunity to connect molecular quantum dynamics at cryogenic temperatures with the chemical inventories observed in space.

\section*{Conflict of Interest Statement}

The authors declare that the research was conducted in the absence of any commercial or financial relationships that could be construed as a potential conflict of interest.

\section*{Author Contributions}

V.L. and M.S-N. outlined the project, performed the laboratory experiments, collected the data, and carried out the theoretical calculations. V.L. analyzed the laboratory data. V.L and M.S.-N. wrote the manuscript and included feedback and comments from all the other authors. V.M.R., I.J.-S., and P.C. coordinated the project.

\section*{Funding}
We gratefully acknowledge the Max Planck Society for the financial support. V.M.R., M.S-N. and I.J-S. acknowledge funding from the grant No. PID2022-136814NB-I00 by the Spanish Ministry of Science, Innovation and Universities/State Agency of Research MICIU/AEI/10.13039/501100011033 and by ERDF, UE. V.M.R. also acknowledges support from the grant number RYC2020-029387-I funded by MICIU/AEI/10.13039/501100011033 and by "ESF, Investing in your future", and from the Consejo Superior de Investigaciones Cient{\'i}ficas (CSIC) and the Centro de Astrobiolog{\'i}a (CAB) through the project 20225AT015 (Proyectos intramurales especiales del CSIC). I.J.-S. also acknowledges support by ERC grant OPENS, GA No. 101125858, funded by the European Union. Views and opinions expressed are however those of the author(s) only and do not necessarily reflect those of the European Union or the European Research Council Executive Agency. Neither the European Union nor the granting authority can be held responsible for them. M.S-N. I.J-S. acknowledge funding from Consejo Superior de Investigaciones Científicas (CSIC) through project i-LINK23017 SENTINEL. M.S.N. also acknowledges a Juan de la Cierva Postdoctoral Fellowship, project JDC2022-048934-I, funded by MCIN/AEI/10.13039/501100011033 and by the European Union “NextGenerationEU/PRTR”.

\section*{Acknowledgments}
Mitsunori Araki is acknowledged for useful insights regarding the theoretical calculations.

\bibliographystyle{Frontiers-Harvard}

\bibliography{tHNSO}

\clearpage

\section{Supplementary Tables and Figures}

\begin{table*}[htbp]
\caption{Calculated equilibrium rotational constants obtained at CCSD[T] level of theory and with different basis sets. All values are in MHz.\\
$^a$Root Mean Square Error defined as \[ \text{RMSE} =  \sqrt{\frac{1}{n}\sum_{i=1}^{n}(\Delta_i)^2}\] where $n$ corresponds to three principal axes of inertia, and $\Delta$ denotes the difference between the experimental and theoretical rotational constants for each axis.\\
}
\label{table:2} \centering
\begin{tabular}{lr@{.}lr@{.}lr@{.}lr@{.}lr@{.}lr@{.}lr@{.}l}
\hline\hline   
\noalign{\smallskip}
Parameter & \multicolumn{2}{c}{cc-pV(T+d)Z} & \multicolumn{2}{c}{cc-pV(Q+d)Z} & \multicolumn{2}{c}{cc-pV(5+d)Z} & \multicolumn{2}{c}{cc-pwCVTZ} & \multicolumn{2}{c}{cc-pwCVQZ}  \\
\hline
\noalign{\smallskip}
   $A_e$       &     50276&193     &  50450&393        &  50758&717  &  50559&954  &  50758&761\\
   $B_e$       &      9918&615     &   9971&256        &  10029&164  &   9945&199  &   9999&057\\
   $C_e$       &      8284&273     &   8325&721        &   8374&490  &   8310&506  &   8353&488\\
\noalign{\smallskip}
\hline
\noalign{\smallskip}
RMSE$^a$  &  91&36  &  21&82  & 199&77 &  76&23  & 194&54 \\
\noalign{\smallskip}

\hline
\end{tabular}
\end{table*}

\vspace{2cm}

\begin{figure*}[htbp]
    \centering
    \includegraphics[scale=0.3]{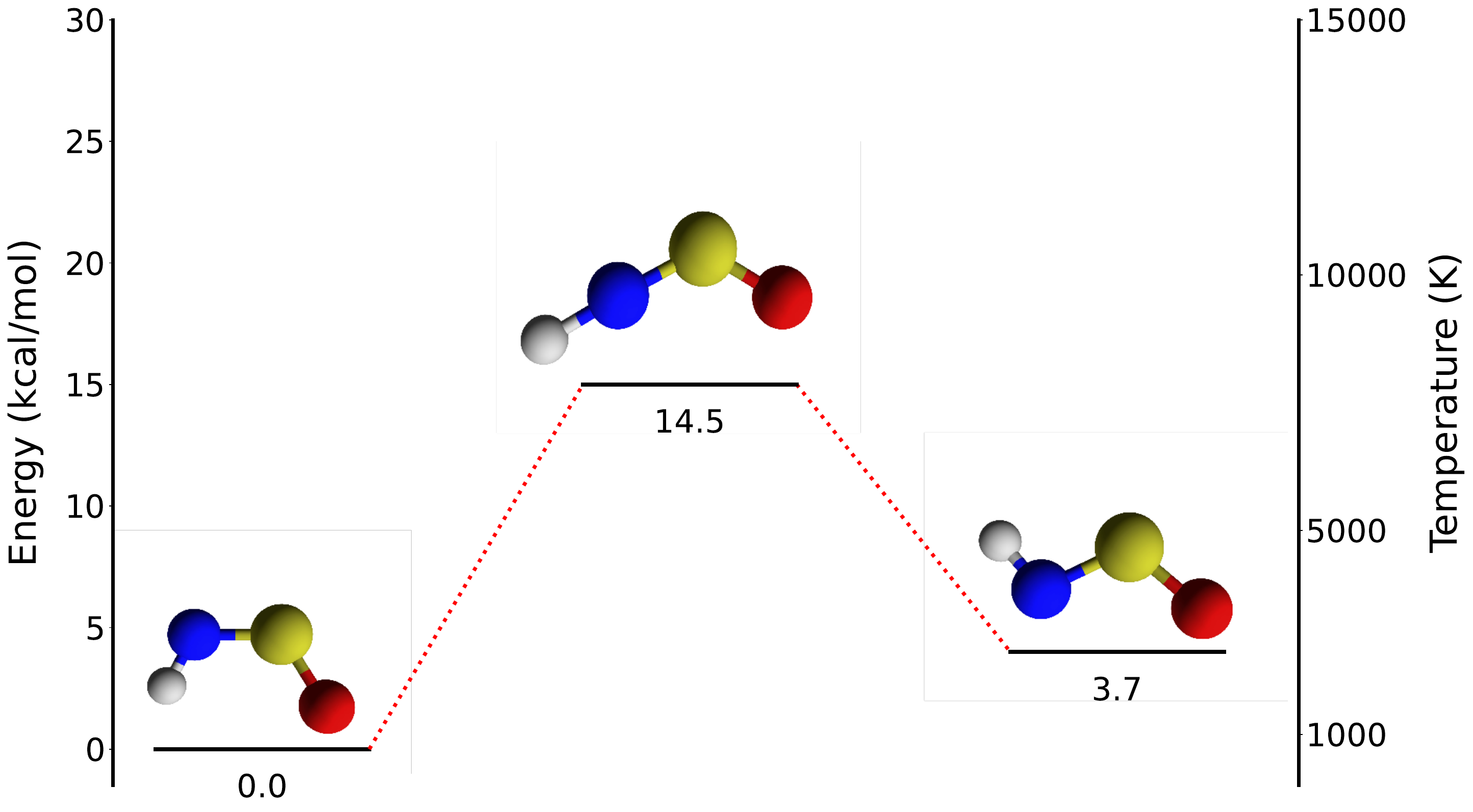}
    \caption{Structures and relative energy levels of $cis$-HNSO (\textit{left}, as the ground state), $trans$-HNSO (\textit{right}, at 3.7\,kcal/mol), and the transition state (\textit{middle}, at 14.5\,kcal/mol); energy values were derived by \cite{Kumar2017}. Red atom corresponds to oxygen, yellow to sulfur, blue to nitrogen, and grey to hydrogen.} 
    \label{fig:isomers}
\end{figure*}

\end{document}